\documentclass[12pt,twoside]{article}

\input{psfig}

\newlength{\dinwidth}
\newlength{\dinmargin}
\newcommand{\beq}[1]{\begin{equation}\label{#1}}
\newcommand{\eeq}{\end{equation}}
 
\newcommand{\beqar}[1]{\begin{eqnarray}\label{#1}}
\newcommand{\eeqar}{\end{eqnarray}}

\newcommand{\gev}{GeV$^2$}
 
\newcommand{\Gl}[1]{Eq.~(\ref{#1})}
\newcommand{\Ab}[1]{Fig.~\ref{#1}}
\newcommand{\Ta}[1]{Table~\ref{#1}}

\newcommand{\ga}{\gamma}
\newcommand{\de}{\delta}
\newcommand{\De}{\Delta}

\newcommand{\si}{\sigma}
\newcommand{\Si}{\Sigma}

\newcommand{\asmz}{\alpha_s(M_Z^2)}

\newcommand{\msbar}{\mbox{$\overline{\rm{MS}}$}\ }

\begin{document}

\begin{flushright}
\tt{NIKHEF/97-028} 
\end{flushright}
\vspace{1.0cm}
\begin{center}
{\Large \bf 
QCD fits to ZEUS and fixed target structure function data}\\
\vspace{0.8cm}
M.A.J. Botje\footnote{Presented on behalf of the ZEUS collaboration
at the 5th International Workshop on Deep Inelastic Scattering
and QCD,
Chicago, IL, April 14--18, 1997}\\
NIKHEF,
PO Box 41882, 1009DB Amsterdam, the Netherlands\\
\vspace{1cm}
\end{center}

\begin{abstract}
\noindent
Preliminary results are presented on the gluon density at low $x$
obtained from a QCD analysis of ZEUS 1994 $F_2$ structure function data
combined with those from NMC. Also given are estimates of the
experimental error on the $e^+p$ NC Born cross section at large $x$ and
$Q^2$. This estimate is obtained from propagation of the statistical
and systematic errors on fixed target structure functions.
\end{abstract}
 
\section*{\centering I\lowercase{ntroduction}}

In perturbative QCD the scaling violations of the $F_2$ structure
functions are caused by gluon bremsstrahlung from quarks and quark
pair creation from gluons. In the low $x$ domain accessible at HERA
the latter process dominates the scaling violations. In this report
I present preliminary results on the gluon momentum density
extracted from a NLO QCD anlysis of $F_2$ structure functions measured by
ZEUS in 1994.

With the present integrated luminosity of about 20 pb$^{-1}$ at HERA
the region of large $x$ ($\sim0.5$) and $Q^2$ ($\sim10^4$ \gev) becomes
accessible. Both ZEUS~\cite{mbb:zhix} and H1~\cite{mbb:h1hix}
have observed an excess of events compared to the standard model
predictions in this hitherto unexplored region. The uncertainty in
these predictions is dominated by that on the parton
distributions in the proton. I will summarise the results of a
QCD analysis by ZEUS of
fixed target structure function data at high $x$ ($>0.1$)
yielding an estimate of the experimental error on the NC Born
cross section of $e^+p$ deep inelastic scattering at HERA.
 
\section*{\centering
QCD \lowercase{analysis of} ZEUS 1994 \lowercase{data}}

The data used in the fit were the ZEUS 1994 nominal vertex
data~\cite{mbb:zeusnv} which cover a kinematic range of
$6.3 \times 10^{-5} < x < 0.5$ and $3.5 < Q^2 < 5000$~\gev\
together with the low $Q^2$ measurements (shifted vertex,
ISR)~\cite{mbb:zeussv}. The latter datasets extend
the kinematic range down to $x = 3.5\times10^{-5}$ and $Q^2 = 1.5$~\gev\
albeit with larger statistical and systematic errors.
NMC data on $F_2^p$ and $F_2^d$~\cite{mbb:nmc} constrain the fit
at high $x$. To remove possible contributions from higher twist
effects at large $x$ the NMC data below $Q^2 = 4$ \gev\ were discarded.

\begin{figure}[b!]
\centerline{\psfig{file=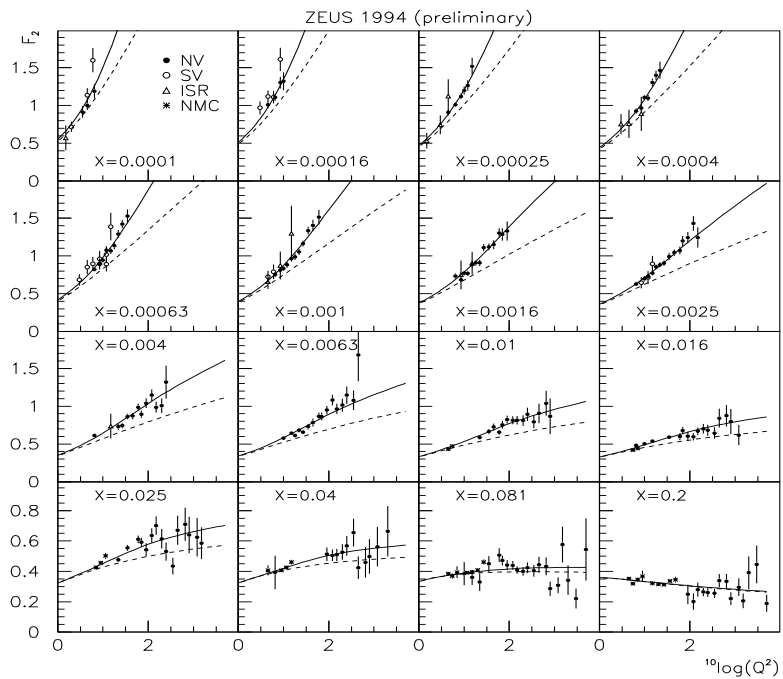,height=10cm,width=12cm}}
\caption{
%{\footnotesize 
The $F_2^p$ structure function
versus $Q^2$ for fixed values of $x$. The errors shown are the 
statistical and systematic errors added in quadradure. The
solid curves correspond to the NLO QCD fit described in the text.
The dashed curves show $F_2$ without the contribution from charm.
%}
}
\label{mbb:fig1}
\end{figure}

At the input scale $Q^2_0 = 7$ GeV$^2$ the gluon distribution ($xg$),
the singlet quark distribution ($x\Si$) and the difference of
up and down quarks in the proton ($x\De_{ud}$) were
parametrised as
\begin{eqnarray} \label{mbb:param}
xg(x,Q_0^2) & = & A_g x^{\de_g}(1-x)^{\eta_g} 
(1 + \gamma_g x) \nonumber \\
x\Sigma(x,Q_0^2) & = &  
A_s x^{\de_s}(1-x)^{\eta_s}(1+\varepsilon_s \sqrt{x}
 + \gamma_s x) \\
x\Delta_{ud}(x,Q_0^2) & = & 
A_{ns} x^{\delta_{ns}}(1-x)^{\eta_{ns}} 
  \nonumber 
\end{eqnarray}
The strange quark distribution was assumed to be 20\% of the sea
at $Q^2 = 4$~\gev~\cite{mbb:sfac}. The sea quark
density was otained by subtracting the valence (taken from MRSD$_-^\prime$)
from the singlet distribution. The gluon normalisation, $A_g$, was fixed by the
momentum sumrule. 
The input value for the strong coupling constant was set to
the result of ref.~\cite{mbb:marcv}: $\asmz =  0.113$.

The input parton distributions were evolved in NLO in the \msbar
scheme with $f = 3$ light flavours. The charm contribution to the
$F_2$ structure function was calculated in NLO from the evolved
distributions as described in~\cite{mbb:f2c} with the charm mass set to
1.5~GeV. Contributions from bottom are estimated to be small and were
neglected.

\begin{figure}[b!]
\centerline{\psfig{file=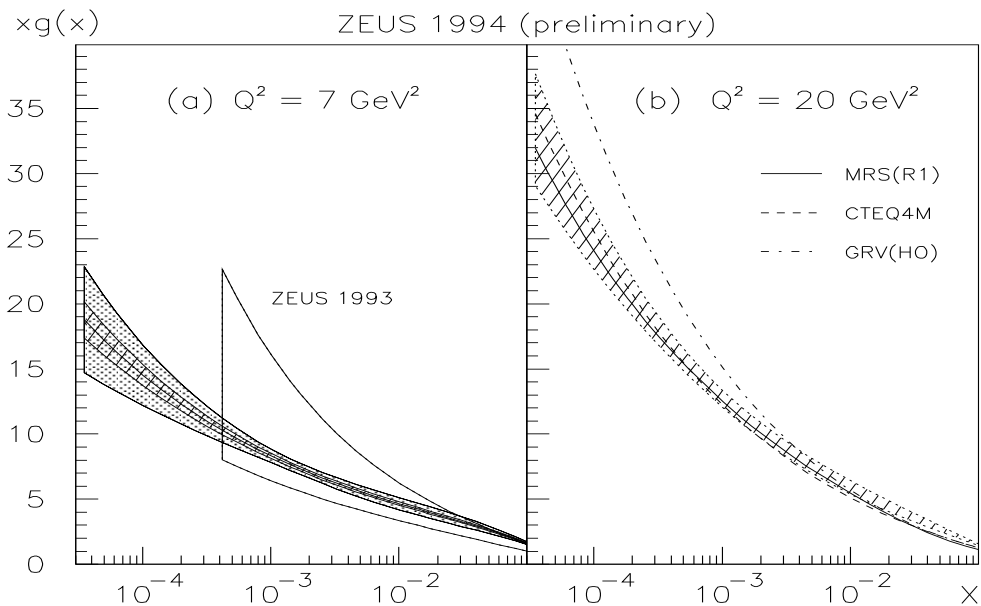,height=10cm,width=20cm}}
\caption{
%{\footnotesize(a) 
The gluon distribution as function of $x$ for a
fixed value of $Q^2 = 7$ \gev. The inner (outer) shaded band
corresponds to the statistical (total) error. Also shown is
the gluon distribution obtained from the ZEUS 1993 analysis.
(b) The gluon distribution from this analysis (shaded band)
at $Q^2 = 20$ \gev\ compared to those from MRS(R1), CTEQ4M
and GRV(HO).
%}
}
\label{mbb:fig2}
\end{figure}

In addition to the 11 parameters describing the parton distributions
one (two) normalisation parameters for the ZEUS (NMC) data
were left free in the fit. 
In the computation of the
$\chi^2$ only statistical errors were taken into account.
For each dataset the quantity
$ [(N-1)/\De N]^2$ was added where $\De N$ is the 
quoted normalisation error of the dataset. 

The fit yielded a good description of the data as shown in \Ab{mbb:fig1}.
Adding the statistical and systematic errors in quadrature the
$\chi^2$ = 463 for 408 
datapoints and 13 free parameters.
The fitted normalisation parameters were 97.4\% for ZEUS and (99.3\%,98.9\%)
for the NMC 90~GeV and 280~GeV datasets respectively which is well within
the quoted normalisation errors.

\Ab{mbb:fig2}a shows the gluon momentum denstity obtained from the
fit at the input scale $Q^2_0 = 7$ \gev. The inner (outer) shaded band
indicate the statistical and the statistical and systematic error
added in quadrature. It is seen that at the lowest value of $x = 3 \times 10^{-5}$
the gluon is determined with an accuracy of about~20\%. At 
$x = 4 \times 10^{-4}$ the total error is $\sim$12\% which is
a large improvement compared to the $\sim$50\% error on the gluon
distribution obtained from the QCD analysis of the ZEUS 1993 $F_2$
data~\cite{mbb:gl93}. At the input scale the momentum fraction carried
by quarks (gluons) was found to be 0.555 (0.445). The fitted values of
the parameters describing
the low $x$ behaviour of the quark singlet and the gluon distributions
are $\de_s = -0.23 \pm 0.01 \pm 0.02$ and
$\de_g = -0.24 \pm 0.02 \pm 0.05$ where the first error is statistical
and the second systematic.

The gluon distribution evolved to
$Q^2 = 20$ \gev\ is shown as the shaded (total) error band in
\Ab{mbb:fig2}b. Also plotted are the gluon densities from the recent
parton distribution sets MRS(R1)~\cite{mbb:mrsr1}, CTEQ4M~\cite{mbb:cteq4m}
and from GRV(HO)~\cite{mbb:grvho}. Whereas the agreement with MRS and
CTEQ at low $x$ is excellent, the steep gluon obtained from the dynamical
evolution by GRV is inconsistent with the result of this analysis.

\section*{\centering E\lowercase{xtrapolation of high \boldmath{$x$}
structure functions to 
large} \boldmath{$Q^2$}}

In this section we describe a QCD analysis of high $x$ structure
function data. The aim of this analysis is to estimate the experimental
uncertainty on the $ep$ neutral current (NC) cross-section at large $x \sim
0.5$ and $Q^2 \sim 3 \times 10^4$. This error estimate is 
used in~\cite{mbb:zhix} to
judge the significance of a possible excess of events observed at HERA
in this region~\cite{mbb:zhix,mbb:h1hix}.

The data used in the fit were the proton and deuteron $F_2$  
data from SLAC~\cite{mbb:slac}, 
BCDMS~\cite{mbb:bcdms} and NMC~\cite{mbb:newnmc}
together with $xF_3^{\nu N}$ from CCFR~\cite{mbb:ccfr}. 
Since we are only interested in the high $x$ domain, 
data below $x = 0.1$ were discarded. Also applied were the
cuts $Q^2 > 4$ GeV$^2$ and $W^2 > 10$ GeV$^2$ to remove target 
mass and higher twist effects. 
$F_2^d$ data above $x = 0.7$ were
discarded to eliminate possible contributions from Fermi motion in
deuterium.

The QCD analysis of these data is similar to that presented in the
previous section except (i)~the charm (bottom) quark distribution
was generated dynamically from the threshold $Q^2_{c(b)} = 4$ (25)
\gev\ (ii)~no momentum sumrule was imposed (iii)~all normalisations
were kept fixed to unity and (iv)~the valence
quark distribution, constrained by the CCFR $xF_3$ data, was left
free in the fit with a functional form identical to that of
$x\De_{ud}$ in \Gl{mbb:param}. From the evolved quark distributions
the $F_2$, $F_L$ and $xF_3$ structure functions in $ep$ scattering
were calculated in NLO including contributions
from $Z_0$ exchange and $\ga Z$ interference~\cite{mbb:fep}.

The experimental systematic errors were propagated to the covariance
matrix of the fitted parameters using the technique described 
in~\cite{mbb:pascaud}. In total 24 independent sources of systematic
error were included taking properly into account the correlations
between the systematic errors of the NMC datasets. 

It is convenient to express the NC cross section  for
e$^\pm$p scattering as
\begin{equation} \label{mbb:rxsec}
\tilde{\sigma}^{\pm} \equiv
\frac {xQ^4} {2\pi\alpha^2 Y_+}
\frac {d^2 \sigma_{NC}^{\pm}}{dx dQ^2} =
F_2 - \frac{y^2}{Y_+} F_L \mp \frac{Y_-}{Y_+} xF_3
\end{equation}
where $y = Q^2/xs$,
$s \approx 4E_eE_p$ is the square of the $ep$ centre of mass energy and
$Y_{\pm} = 1 \pm (1-y)^2$. 

In \Ab{mbb:fig3}a we show the cross section for $e^+p$ scattering
at HERA energies ($E_e = 27.5$ $E_p = 820$ GeV) as a function
of $Q^2$ for fixed values of $x$. 
%The fixed target $F_2$ data
%shown in this figure are converted to $\tilde{\si}^+$ by making a small
%correction for $F_L$. 
The strong decrease in the cross section
above $Q^2 \sim 5000$ \gev\ is due to the contribution from
$xF_3$. 
At high $x \sim 0.5$ and $Q^2 \sim 3 \times 10^4$ GeV$^2$
the error on $\tilde{\sigma}^+$ is estimated to be about 9\%,
see \Ta{mbb:tab1}. The estimated errors include (small) contributions
from an assumed uncertainty of $\Delta \alpha_s (M_Z^2) = 0.005$
and an error of 50\% on the strange quark content of the proton.

\begin{table}[b]
\label{mbb:tab1}
\begin{center}
\begin{tabular}{ccccccccccc}
  $Q^2$ (\gev) & $x \rightarrow$  & 0.1 & 0.2 & 0.3 & 0.4 & 0.5 & 0.6 &
0.7 &  0.8 & 0.9 \\
\hline
1000  & & 4 & 4 & 4 & 4 & 4 & 5 & 8 & 15 & 31 \\
30000 & &   &   &   & 11 & 9 & 8 & 9 & 14 & 30 \\
\hline
\end{tabular}
\end{center}
\caption{The relative uncertainty in $\si^+_{NC}$ in percent.}     
\end{table}

In \Ab{mbb:fig3}b the cross sections obtained from this analysis
are compared to those calculated from the parton distribution
sets MRSA~\cite{mbb:mrsa} and CTEQ3M~\cite{mbb:cteq3m}. It is seen that the
differences are much smaller than the experimental errors: comparison
of results obtained from different parton distribution sets do not 
yield a reliable estimate of the uncertainty in the cross sections.

\begin{figure}[t!]
\centerline{\psfig{file=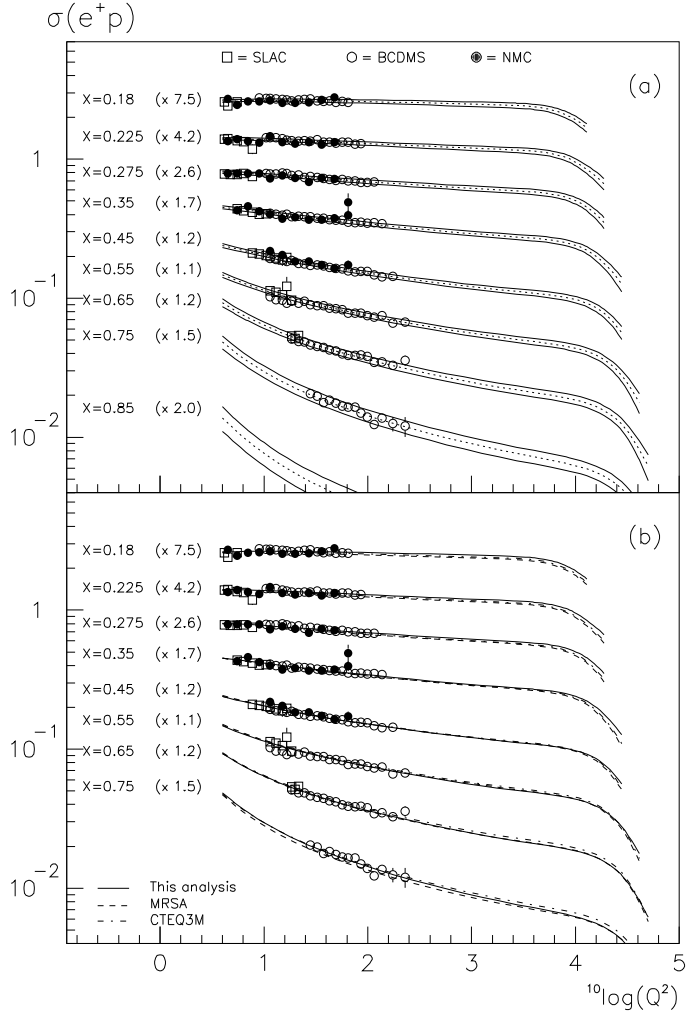,height=15cm}}
\caption{
%{\footnotesize
(a) The cross-section $\tilde{\sigma}$ (see text)
for $e^+p$ scattering at Hera energies.
(b) The result of this analysis compared to cross sections calculated
with the parton distribution sets MRSA and CTEQ3M.
%}
}
\label{mbb:fig3}
\end{figure}


\begin{thebibliography} {99}

\bibitem{mbb:zhix}
  ZEUS Collab.,J. Breitweg et al., {\it Z. Phys.} {\bf C74}, 207 (1997).
\bibitem{mbb:h1hix}
  H1 Collab., C. Adloff et al., {\it Z. Phys.} {\bf C74}, 191 (1997).
\bibitem{mbb:zeusnv}
  ZEUS Collab.,M. Derrick et al., {\it Z. Phys.} {\bf C72}, 399 (1996).
\bibitem{mbb:zeussv}
  ZEUS Collab.,M. Derrick et al., {\it Z. Phys.} {\bf C69}, 607 (1996).
\bibitem{mbb:nmc}
  NMC, P. Amaudruz et al., {\it Phys. Lett.} {\bf B295}, 159 (1992).
\bibitem{mbb:sfac}
  CCFR Collab.,C. Foudas et al., {\it Phys. Rev. Lett.} {\bf 64}, 1207
  (1994); S. A. Rabinowitz et al., {\it Phys. Rev. Lett.} {\bf 70},
  134 (1993).
\bibitem{mbb:marcv}
  M. Virchaux and A. Milsztajn, {\it Phys. Lett.} {\bf B274}, 221 (1992).
\bibitem{mbb:f2c}
  S. Riemersma et al., {\it Phys. Lett.} {\bf B347}, 143 (1995) and
  references therein.
\bibitem{mbb:gl93}
  ZEUS Collab.,M. Derrick et al., {\it Phys. Lett.} {\bf B345}, 576 (1996).
\bibitem{mbb:mrsr1}
  A.D. Martin, R.G. Roberts and W.J. Stirling, DTP/96/44,
  RAL-TR-96-037, hep-ph/0606345 (1996).
\bibitem{mbb:cteq4m}
  H.L. Lai et al., MSUHEP-60426, CTEQ-604, hep-ph/9606399 (1996).
\bibitem{mbb:grvho}
  M. Gl\"{u}ck, E. Reya and A. Vogt, {\it Z. Phys.} {\bf C67}, 433 (1995).
\bibitem{mbb:slac}
  L.W. Whitlow et al., {\it Phys. Lett.} {\bf B282}, 475 (1992).
\bibitem{mbb:bcdms}
  BCDMS, A.C. Benvenuti et al., {\it Phys. Lett.} {\bf B223}, 485 (1989)
  and {\it Phys. Lett.} {\bf B237}, 592 (1990).
\bibitem{mbb:newnmc}
  NMC, M. Arneodo et al., {\it Nucl. Phys.} {\bf B483}, 3 (1997).
\bibitem{mbb:ccfr}
  CCFR Collab., P. Z. Quintas et al., {\it Phys. Rev. Lett.} {\bf 71}
  1307 (1993).
\bibitem{mbb:fep}
  G. Ingelman and R. R\"{u}ckl, {\it Z. Phys.} {\bf C44} 291 (1989); \\
  J. Bl\"{u}mlein et al., {\it Z. Phys.} {\bf C45} 501 (1990).
\bibitem{mbb:pascaud}
  C. Pascaud and F. Zomer, preprint LAL 95-05 (1995).
\bibitem{mbb:mrsa}
  A.D. Martin, R.G. Roberts and W.J. Stirling, {\it Phys. Rev.} 
 {\bf D50} 6734 (1994).
\bibitem{mbb:cteq3m}
  H.L. Lai at al., {\it Phys. Rev.} {\bf D51}, 4763 (1995).

\end{thebibliography}
\end{document}